\documentclass[usenatbib]{mn2e}
\usepackage{psfig,morefloats,url,graphicx}
\usepackage[multiple]{footmisc}

\newcommand{\hii}{{\rm H\,{\footnotesize II}}}
\footnotesize
\newdimen\digitwidth    
\setbox0=\hbox{\rm0}
\digitwidth=\wd0
\catcode`!=\active
\def!{\kern\digitwidth}
\normalsize

\title[Pulsar spectral turnovers]{On gigahertz spectral turnovers in pulsars}
  
\author[K. Rajwade et al.]
  {K.~Rajwade\thanks{Email: kmrajwade@mix.wvu.edu},
  D.~R.~Lorimer, L.~D.~Anderson \\
  Department of Physics and Astronomy, West Virginia University, Morgantown, WV 26506, USA}
  
\date{\today}

\def\LaTeX{L\kern-.36em\raise.3ex\hbox{a}\kern-.15em
    T\kern-.1667em\lower.7ex\hbox{E}\kern-.125emX}

\begin{document}

\label{firstpage}

\maketitle

\begin{abstract}
 Pulsars are known to emit non-thermal radio emission that is
 generally a power-law function of frequency. In some cases, a
 turnover is seen at frequencies around 100~MHz. Kijak et al.~have
 reported the presence of a new class of ``Gigahertz Peaked Spectrum''
 (GPS) pulsars that show spectral turnovers at frequencies around
 1~GHz. We apply a model based on free-free thermal absorption to
 explain these turnovers in terms of surrounding material such as the
 dense environments found in \hii\ regions, Pulsar Wind Nebulae (PWNe),
 or in cold, partially ionized molecular clouds. We show that the turnover frequency depends on the
 electron temperature of the environment close to the pulsar, as well
 as the emission measure along the line of sight. We fitted this model
 to the radio fluxes of known GPS pulsars and show that it can
 replicate the GHz turnover. From the thermal absorption model, we
 demonstrate that normal pulsars would exhibit a GPS-like behaviour if
 they were in a dense environment. We discuss the application of this
 model in the context of determining the population of neutron stars
 within the central parsec of the Galaxy. We show that a
 non-negligible fraction of this population might exhibit
 high-frequency spectral turnovers, which has implications on the
 detectability of these sources in the Galactic centre.

\end{abstract}


\section{Introduction}

Radio flux measurements from pulsars have revealed a wealth of
information about the underlying physical processes involved in
coherent radio emission (for a review, see Graham-Smith 2003).  A
knowledge of the spectral behaviour of pulsars is also important for
population studies that seek to constrain the luminosity function of
the underlying population (e.g., Bates et al.~2014). Studies in the
past have shown that the flux density, $S$, as a function of
frequency, $\nu$, for a pulsar can be described by a simple power law
$S \propto \nu^{\alpha}$, with a spectral index $\alpha$~(e.g.,
Lorimer et al.~1995). Although the observed spectra are found to have
a mean value of $\alpha$ around --1.6 (Maron et al.~2000), population
models suggest that the underlying population is more consistent with
a normal distribution of spectral indices with the mean value around
--1.4 (Bates et al.~2013). A small fraction of such sources ($\sim$
10$\%$) show a broken power-law behaviour, with $\alpha$ of $\sim$
--0.9 and $\sim$ --2.2 (Maron et al.~2000). At low frequencies,
synchrotron self absorption becomes dominant in the pulsar
magnetosphere and the spectrum tends to show a turnover
(Sieber~1973). Such turnovers have been detected for many pulsars
around $\sim 100$~MHz. An excellent example of such a spectrum can be
found for PSR~B0329+54 in Kramer et al.~(2003).

In 2007, a new class of pulsars was proposed. As described by Kijak et
al.~(2007, 2011; hereafter K07 and K11), these ``Gigahertz Peaked
Spectrum'' (GPS) pulsars show a spectral turnover at frequencies $\sim
1$~GHz. To date, eleven GPS pulsars have been reported of which two are magnetars (K07, K11, Dembska
et al.~2014, Lewandowski et al~2015).  These authors have found a correlation between the spectral shape of pulsars showing such behaviour and the
environment around the pulsar. The strongest argument for
environmental origins of high-frequency turnover came from the
observations of the binary pulsar B1259--63 (Kijak et al. 2011; hereafter K11a). This
pulsar exhibits GPS-like behaviour when it is close to its companion
Be star LS~2883 and shows a single power-law spectrum when it is far
from the Be star. In another study, Kijak et al.~(2013) obtained
spectra for two magnetars that show GPS-like behaviour. Both of the
magnetars are associated with supernova remnants (SNRs).  The presence
of these dense, ionized regions around the pulsars suggests that
free-free absorption by the surrounding ionized gas could be
responsible for high-frequency turnovers. The authors concluded that
pulsars located within ionized environments such as SNRs,
\hii\ regions, and PWNe that have high electron densities and
emission measures should invariably have high-frequency spectral
turnovers.

Motivated by these ideas, a very recent study by Lewandowski et
al.~(2015) applied a similar model to the one presented here. They show that the absorption is dominant in moderately cold
    plasma ($T_{e}\sim 5000-8000$\,K) with heightened electron
    densities (above $\sim 1000~{\rm cm^{-3}}~{\rm pc}$).  They use
their model to show that the rapid variations in the spectrum of radio
magnetar SGR 1745$-$2900 in the Galactic centre can be explained by
free-free thermal absorption of the radio emission by ejecta
surrounding the magnetar.  Using simulations, they were able to show
that pulsars can exhibit GPS behaviour that can be explained by the
model. This is compelling evidence of the dependence of pulsar spectra
on their surroundings.

Thermal absorption by free electrons in the vicinity of the
  GPS pulsars may explain their spectral turnovers as proposed by Kijak et al~(2011) and later by Kijak et al~(2013). Radio emission
from pulsars is known to have a steep spectrum that is believed to be
due to a non-thermal emission process consisting of pair production in
the magnetosphere (see, e.g., Contopoulos \& Spitkovski 2006; Melrose
et al.~2014).  The characteristics of this emission are similar to
those of radio emission observed from SNRs.  In this paper, we modeled
the emission from pulsars based on free-free absorption, which has
previously been used to explain the radio emission from SNRs.  A
similar approach was taken by Lewandowski et al.~(2015).  In contrast
to these authors, we have fitted this thermal absorption model to the
six known GPS pulsars to constrain the Emission Measure (EM) along the
line of sight based on known electron temperatures of the environment
surrounding them. In addition, we consider multiple
    sources of absorbers and try to obtain the most suitable physical
    conditions for the observed spectrum. We also looked at one bright pulsar that does not show GPS
behaviour. Using known parameters of EM and electron temperatures for
PWNe, we simulated the spectrum of this source to that
    it can exhibit GPS behaviour and discuss about the pulsar's 
    low-frequency turnover. The model, procedure and results of our
fitting are explained in section 2. In section 3, we discuss various
implications of our results, and in particular highlight the
importance of the model on the detectability of pulsars in the
Galactic centre.
     
\section{Model}

\subsection{Theory}

Starting from the fundamental equation of radiative transfer (see,
e.g., Burke \& Smith 2014), we considered 2 scenarios: (i) the
pulsar lies within the PWN/SNR or beyond an \hii\ region; (ii) the pulsar lies beyond a cold, partially ionized molecular cloud. In either case, the total
measured flux
\begin{equation}
S_{{\rm obs,\nu}} = S_{{\rm psr,\nu}}e^{-\tau_\nu} + S_{{\rm reg,\nu}}
(1 - e^{-\tau}),
\end{equation}
where $S_{{\rm psr,\nu}}$ is the pulsar's intrinsic flux,
$S_{{\rm reg,\nu}}$ is the flux of the intervening region (PWN
or \hii\ region) and $\tau_\nu$ is the optical depth at frequency $\nu$ for this line of sight.  We
can ignore the $S_{{\rm reg,\nu}}$ term because it is a continuum
emission that adds to the sky background. Assuming $\tau_\nu
\sim \nu^{-2.1}$ for free-free absorption and assuming $\tau_\nu \ll 1$ for the frequencies of interest (Mezger \& Henderson, 1967) we get
\begin{equation}
\label{eq:model}
S_{{\rm obs,\nu}} = S_{\rm ref}\left(\frac{\nu}{\nu_{\rm
    ref}}\right)^{\alpha} \exp \left[-\tau_{\rm ref}\left(\frac{\nu}{\nu_{\rm ref}}\right)^{-2.1}\right].
\end{equation}
Here $S_{\rm ref}$ is some reference flux density measured at a reference frequency $\nu_{\rm ref}$, $\alpha$
is the spectral index and $\tau_{\rm ref}$ is the optical depth at the
reference frequency.  This is similar to the model previously
developed to fit spectra of SNRs (see, e.g., Dulk \& Slee~1975).

To obtain the optical depth at a given frequency we used the
expression given in Altenhoff et al.~(1960) and Mezger \& Henderson
(1967) where we can assume that the medium is optically thin
($\tau \ll 1$) at high frequencies.
Under this assumption, the optical depth
\begin{equation}
\tau = 0.082 a \left(\frac{\nu}{\rm GHz}\right)^{-2.1} 
\left(\frac{\rm EM}{{\rm cm}^{-6}~{\rm pc}}\right) 
\left(\frac{T_{e}}{\rm K}\right)^{-1.35},
\label{eq:tau}
\end{equation}
where $a$ is a correction factor of the order unity for
electron temperatures $T_{e} > 20$~K and EM is the emission measure.
Using equations~\ref{eq:model} and~\ref{eq:tau}, we find the spectrum
peaks at a frequency
\begin{equation}
\nu_{\rm peak} = 0.433 \, {\rm GHz} \, (-\alpha)^{-0.476}
\left(\frac{\rm EM}{{\rm cm}^{-6}~{\rm pc}}\right)^{0.476}
\left(\frac{T_e}{\rm K}\right)^{-0.643}.
\label{eq:peak}
\end{equation}

\subsection{Application}
 
We fitted equation~\ref{eq:model} to the flux density spectra for six GPS pulsars reported in K11. The flux data were taken from K07, K11 and Dembska et al~(2014). For PSR~B1054$-$62 and PSR~J1852$-$0635, the published
errors on the fluxes (Dembska et al.~2014) were substantially smaller
than the errors for other pulsars, which were around 20$\%$. We
therefore increased the errors on these two pulsars to make them
comparable with the rest of the sample. We did this fitting
  for two scenarios: one for warm plasma with characteristic
  properties of \hii\ regions or PWNe and one for cold, partially
  ionized clouds.

  For the first scenario, we use characteristic properties of
  PWNe/\hii\ regions, which are known to have $T_e \simeq 5000-10^4$~K for
  \hii\ regions and $T_e \simeq 10^{4}-10^6$~K for PWNe (e.g., Slane et
  al.~2004). The value of $T_{e}$ for the
      PWN/\hii\ region scenario was fixed to 5000~K as Lewandowski et
      al.~(2015) show that the absorption of radio emission takes
      place in plasma with ionized gas filaments with
      $T_{e}\sim$5000~K in SNRs (Sankrit et al.~1998; Koo et
      al.~2007). Such high density plasma also exists in young ``ultra
      compact'' star forming \hii\ regions where electron densities
      are of the order of $\sim 10^{4} \rm cm^{-3}$ (for a review, see
      Churchwell~2002) so it is also possible that a pulsar lying
      beyond an \hii\ region might experience absorption of radio
      emission. Pulsars within PWNe may not exhibit GPS spectra
      because the PWNe plasma distribution is inhomogeneous and there
      may not be a dense filament between us and the pulsar.

  For the second scenario, we use characteristic properties of
  cold, partially ionized clouds, which Dulk \& Slee~1975
    suggest are the most promising absorbers of radio emission from
    SNRs.  For such clouds, we used $T_{e} = 30$\,K. Due to asymmetries
    in the PWN shell, the radio emission may be absorbed by the cold
    clouds instead of the filaments. However, PWN are often found in
    the vicinity of molecular clouds with a high rate of star
    formation.  We kept $\alpha$ and $\tau$ as free parameters as we
believe that there could be a bias in the measured values of $\alpha$
due to the high-frequency turnover behaviour. The value of EM can be
calculated from the derived value of $\tau$ and by assuming a $T_{e}$. For
this case, we assumed 30$\%$ uncertainties in the value of $T_{e}$,
and propagated this uncertainty to derive uncertainties in EM. The
value of $30\%$ is arbitrary in the sense that it only affects the
errors on derived values of EM and not the values themselves. We
select $S_{\rm ref}$ and the corresponding $\tau_{\rm ref}$ from the
highest frequency measurement since the effects of free-free
absorption should be negligible at those frequencies and therefore the
assumptions we made for $\tau_{\rm ref}$ would hold. Using this as our
starting point, we fitted the model to the observed spectra for the
two different electron temperatures scenarios mentioned above using
the Levenberg-Marquardt non-linear least squares algorithm (see, e.g.,
Press et al.~2010) and for each pulsar derived the values for
EM and $\alpha$ given in Table~\ref{tab:der_val}.

    The values of EM we derived in Table
    ~\ref{tab:der_val} have rarely been measured before. Here, we give
    a possible physical explanation of why such high values can
    arise. From the simulations done in Lewandowski et al.~(2015), one
    infers that to observe GPS behaviour, a pulsar needs to be beyond
    a region of ionized gas a fraction of a parsec thick with ambient
    temperatures of the order of a few 1000~K and relatively high free
    electron density. High electron densities of ($\sim 2000-6000~{\rm
      cm^{-3}}$) with relatively cooler temperatures ($\sim
    5000-8000$~K) have been found to exist in dense filaments a
    fraction of a parsec wide around SNRs and PWNs (Sanskrit et
    al.~1998; Koo et al.~2007). For example, if we consider the line
    of sight along the filament found by Koo et al. in SNR G11.2-0.3,
    we can infer an EM contribution due to filament by using the
    values in their paper which is $\sim10^{7}~{\rm cm^{-6}~{\rm
        pc}}$. This is of the same order as our derived values. If we
    assume that the derived EM contributes to the total DM by a
    fraction $a$ then the linear size of the filament
\begin{equation}
d_{\rm filament} = a^2 \left(\frac{\rm DM}{{\rm cm}^{-3}~{\rm pc}}\right)^2 \left(\frac{\rm EM}{{\rm cm}^{-6}~{\rm pc}}\right) \, {\rm pc}.
\label{eq:dim}
\end{equation} 
The values obtained from Equation~\ref{eq:dim} reflect the dimension of the filament along the line of sight. For the absorbing medium,
the mean electron density
\begin{equation}
\langle n_{e} \rangle = \frac{1}{a} \left(\frac{\rm DM}{{\rm cm}^{-3}~{\rm pc}}\right)^{-1} \left(\frac{\rm EM}{{\rm cm}^{-6}~{\rm pc}}\right)
\, {\rm cm}^{-3}.
\end{equation}

    We estimate the electron density and the linear size
    of the absorber by assuming a DM fraction of $50\%$ for the GPS
    sources in this paper. We did the analysis for the absorption
    scenarios considered for the previous analysis. From our results,
    we infer the most likely source of absorption for all the pulsars
    in our sample. The values of electron density, linear size and the
    source of absorption are listed in Table~\ref{tab:params}. The
values we obtained for $\nu_{\rm peak}$ differ from values given K11
simply because K11 calculate the turnover frequency from the
intersection of two linear fits to the spectrum. To quantify the
quality of the fit even further, we obtained the reduced $\chi^2$ by
fitting the pulsar fluxes with a single power law and compared them to
the ones obtained by the model. The values for the power law are
higher by at least a factor of two, suggesting that the model given in
equation \ref{eq:model} is preferred over the power-law model.

\begin{table*}
\begin{tabular}{lc lc lc lc lc lc lc lc}
\hline
PSR & $\alpha$ &$\tau$ &\multicolumn{2}{l} {EM}  & $\nu_{\rm peak}$ & $\chi^{2}$ & $\chi^{2}$\\
    &   & & (10$^{6}$~pc~cm$^{-6}$) & & (GHz)& Model & Power law  \\
\hline
 &  & & 5000K&30K & & & &\\
\hline
B1054$-$62 & --2.8(8)& 0.19(3) &1.1(9)&0.0006(4) &0.6&3.5 & 6.5 \\
J1809$-$1917 & --2.5(4) &0.16(3) &5.2(3) & 0.005(3) & 1.8 & 1.2&7.3 \\
B1822$-$14 & --0.6(1) & 0.005(1)& 0.2(1) &0.0002(1) & 0.7 & 1.7 & 16.8 \\
B1823$-$13 & --0.8(1)&0.003(1)& 0.5(2) &0.0005(2) & 1.0 & 2.3 & 46.5 \\
J1740+1000 &--2.0(1) &0.007(3)&0.8(1)  &0.0008(1) & 0.8 & 12 & 122.8  \\
J1852$-$0635 &--1.1(1) &0.006(2)& 0.7(3)&0.0007(3) & 1.0 & 1.3 & 3.8 \\
\hline
\end{tabular}
\caption{For each pulsar, we list the derived
  values of $\alpha$ and $\tau$ from fitting Equation 2 to the spectra.
  Also listed are the assumed electron temperature and derived EM using
  Equation 3, $\nu_{\rm peak}$ calculated from Equation 4 as well as 
  the reduced $\chi^2$ values from fitting out model versus a simple
  power law. Figures in parentheses represent the
  formal uncertainties in the least significant
  digits.\label{tab:der_val}}
\end{table*}

\begin{figure*}
\centering
{\mbox{\psfig{file=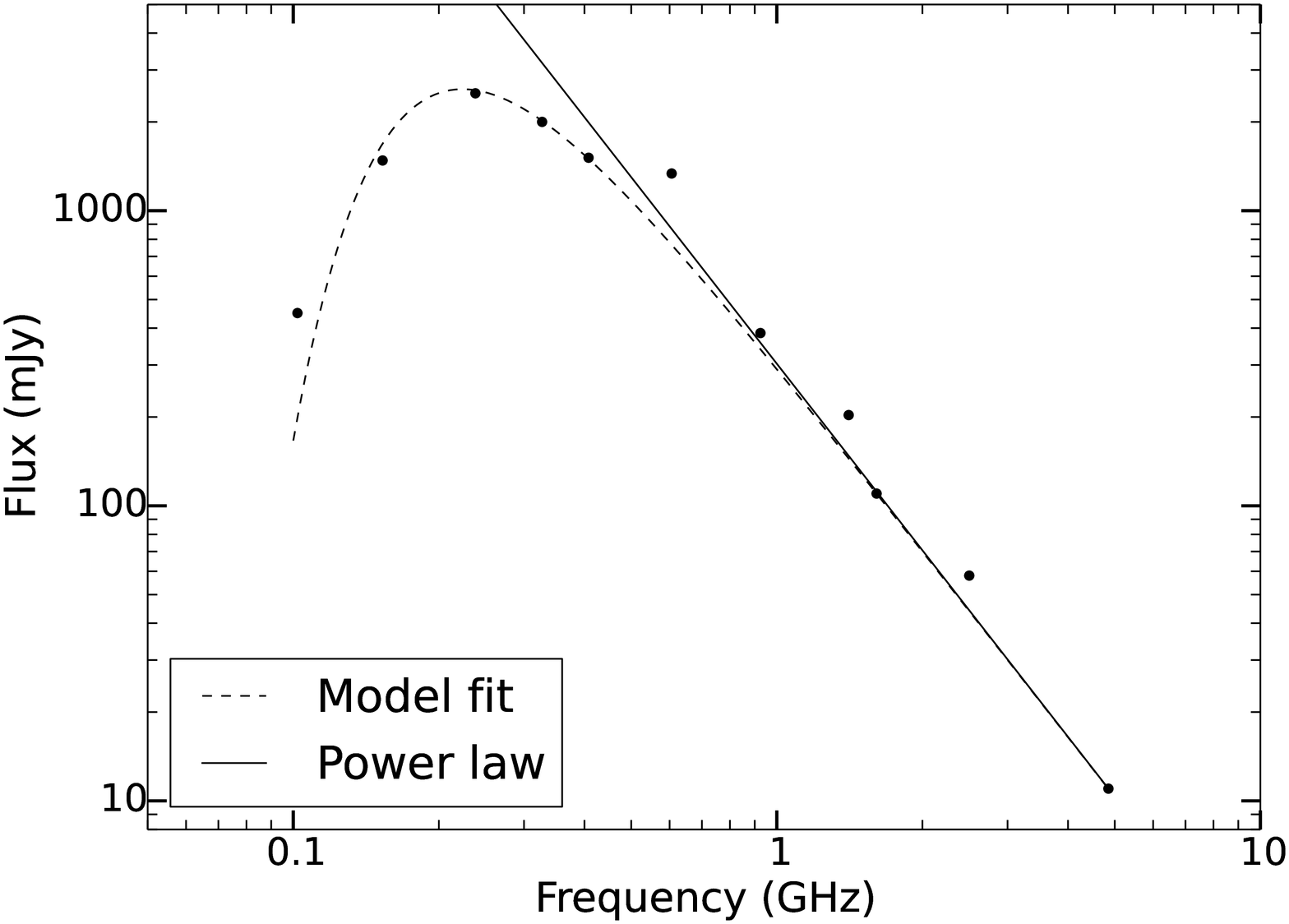,width=12cm,height=8cm}}} 
\caption{Power law spectrum (solid line) and the best fit curve (dashed line) from the model for PSR~B0329+54. The black filled circles are measured flux densities taken
  from literature. The power law has a spectral index $\alpha$ = $-$2.1 taken from the literature. The  model fit has a reduced $\chi^{2}$ value of 1.8. 
\label{fig:simspec}}
\end{figure*}

\begin{figure*}
\psfig{file=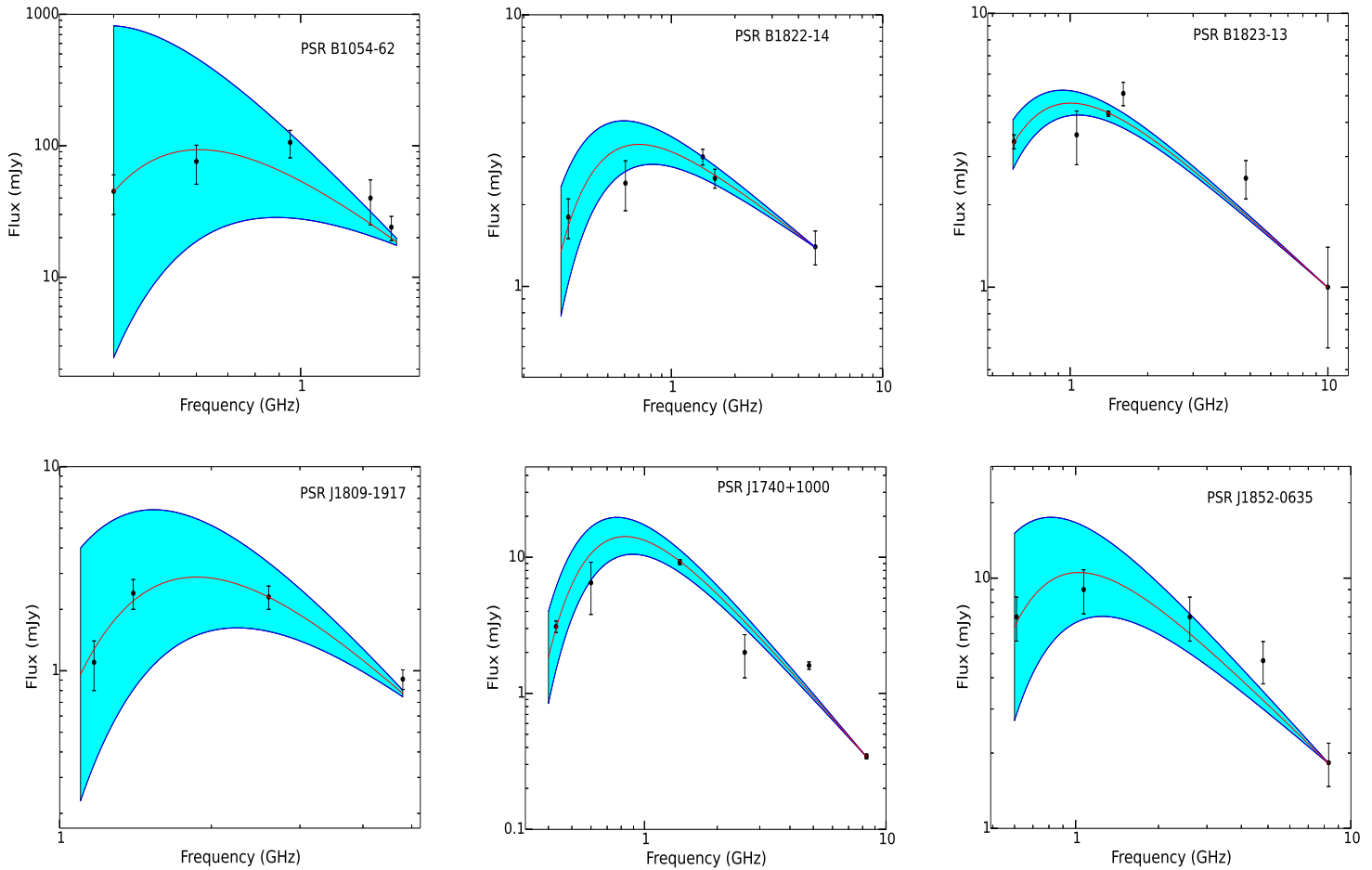,width=20cm,height=13cm}
\caption{Best fit curve along with $\pm 1\sigma$ region (shaded) obtained from
  the thermal absorption model for the set of GPS pulsars. Black circles are measured
  flux densities taken from K07, K11 \& Dembska et al.~(2014). The $\pm 1\sigma$ shaded region is determined by assuming a Gaussian error on the derived parameters and using $\pm 1\sigma$ limits of these parameters to obtain the curves.}
\label{fig:Fit}
\end{figure*}

\begin{table*}
\begin{tabular}{l l l l l l}
\hline
PSR &\multicolumn{2}{l}{Electron density} &\multicolumn{2}{l}{Linear size} & Absorber
\\
& (cm$^{-3}$)& & (pc) & &\\
\hline
&5000~K &30~K &5000~K &30~K &  \\
\hline
B1054$-$62 &3.6$\times$10$^{3}$ &3.6 &0.044 &48 &Cold, partially ionized cloud \\
J1809$-$1917 &5.3$\times$10$^{4}$ &53.5 &0.002 &1.8 &Cold, partially ionized cloud \\
B1822$-$14 &0.9$\times$10$^{3}$ &0.9 &0.2 &195 & Dense, ionized filament \\
B1823$-$13 &4.0$\times$10$^{3}$ &4.2 &0.02 &27.2 &Dense, ionized filament\\
J1740+1000 &6.6$\times$10$^{4}$ &66 &0.0002 &0.2 & Cold, partially ionized cloud \\
J1852$-$0635 &8.0$\times$10$^{3}$ &7.7 &0.01 &11 &Cold, partially ionized cloud \\
\hline
\end{tabular}
\caption{Constrained parameters for all GPS pulsars in our sample by assuming an absorber contribution of 50$\%$ to the total DM of the pulsar. To infer the absorber, we assumed $n_{e}=$100$-$6000~cm$^{-3}$ for a dense, ionized filament and $n_{e}=$1$-$100~cm$^{-3}$ for a cold, partially ionized cloud and also considered the known environment around the pulsar. PSR~B1054$-$62 was an exception (see text for details).
\label{tab:params}}
\end{table*}

The model fits the observed data well, as illustrated by
Fig.~\ref{fig:Fit} and the reduced $\chi^{2}$ values given in
Table~\ref{tab:der_val}. This motivated us to look at pulsars that do
not lie in dense environments and to simulate their spectra to examine if the low-frequency turnover is due to the environment or the pulsar itself. 
For this purpose, we selected
one bright, non-GPS pulsar, PSR~B0329+54, for which
there are reliable estimates of flux densities at different
frequencies and spectral index in literature (Kramer et
al.~2003). Using the model, we tried to fit the spectrum of this
pulsar. The slight turnover seen in the actual data for~PSR B0329+54
is at $\sim$ 200 MHz. We believe that this turnover is
    intrinsic to the pulsar itself. Hence, the fit should give us
    values of EM due to ISM, or it may give some unreasonable values
    which might suggest a different mechanism for the low-frequency
    turnover seen in Fig.~\ref{fig:simspec}. To show this, we fitted
for the EM of PSR B0329+54 by constraining the electron temperature,
which we assume to be 5000~K (i.e., that of the Warm Ionized Medium
(WIM) (Madsen et al.~2006)). The value we obtained was
$5.2\times10^{4}~{\rm cm}^{-6}~{\rm pc}$. If we assume a WIM-dominated
ISM between the pulsar and the observer, we can use the measured
parallax for PSR~B0329+54 (Brisken et al.~2002) and its measured
dispersion measure (DM), the integrated electron density along the
line of sight to find a mean electron density of $\sim 0.024~{\rm
  cm}^{-3}$. Assuming a filling factor of~0.1 (Berkhuijsen and Muller
2008) to account for the clumpiness of the ISM, and knowing the
distance to the pulsar, we derive a value of EM of $\sim 0.052 ~{\rm
  cm}^{-6}~{\rm pc}$. This value of EM is $\sim$6 orders of magnitude
smaller than the one derived from the model which suggests that the
turnover in spectrum of PSR B0329+54 could be due to synchrotron
self-absorption in the pulsar's magnetosphere. Also, if we fix the EM
to $\sim 0.052 ~{\rm cm}^{-6}~{\rm pc}$ and fit for the electron
temperature, the value we obtain is 0.18~K which is nonsensical for a
WIM-dominated ISM.

\section{Discussion}
Long-period pulsars are known to show turnovers in their flux density
spectra at frequencies of $\sim 100$~MHz (Maron et al.~2000). It is
proposed that at such low frequencies, the radio emission becomes
optically thick because of synchrotron self-absorption (O'Dea~1998,
Chevalier~1998). From the thermal absorption model, it is seen that if
the pulsar were to lie in an extremely dense environment, free-free
absorption in the dense region can dominate at frequencies higher than
100~MHz depending on the electron density and electron temperature of
the environment. This results in the pulsar flux being absorbed by the
surrounding material, which manifests itself as a high-frequency
turnover. This study is consistent with the claim that GPS behaviour
does not depend on the DM of the pulsar (K11; Dembska et
al.~2014). All pulsars lying in a high electron density environment
would invariably have high DMs but we measure a higher DM even if the
pulsar is further away from us and not necessarily in a dense
environment.  If this were not true, all pulsars with a high DM would
have shown GPS-like characteristics. It is important to note that
only pulsars where the line of sight traverses through
    such dense filaments might show GPS behaviour.
    We derive large values for the EM for which we
    consider two physical scenarios based on previous studies (Dulk \& Slee, 1975; Lewandowski et al. 2015). Either
    the pulsar flux is absorbed by the high density, ionized filaments
    surrounding the PWNe/SNRs or by the cold, partially ionized clouds along the line of sight. Using this idea, and the fact that
    the absorbers only contribute to a part of the observed DM, we
    calculate the mean electron density of the absorbers. Assuming a
    certain range of values of $n_{e}$ for
    each scenario ($n_{e}\sim$100$-$6000~cm$^{-3}$ for SNR filaments and $n_{e}\sim$1$-$100~cm$^{-3}$ for cold molecular clouds) and considering the known environment of each
    pulsar, we report the most plausible absorber for each pulsar in
    our sample in Table~\ref{tab:params}. For PSR~B1054$-$62, both the
    scenarios give reasonable estimates for the electron densities. We
    believe that a cold absorber is more suitable because, although
    the pulsar lies near an \hii\ region RCW55 (Koribalski et al.~1995),
    it lies at the very edge of the region so there is not enough high
    density, ionized material to produce the observed spectrum. Also,
    the fact that PSR~B1054$-$62 lies in the Carina complex, a
    large molecular complex with high star formation, strengthens our
    claim. The calculations of electron density provides an
    independent estimate of electron densities within the dense clumps
    of the ISM that can be very difficult to obtain by conventional
    methods as most of emission we detect from these sources is
    non-thermal synchrotron emission (Kargaltsev et al.~2007).

The model also can be useful to gain insights into the emission
physics of millisecond pulsars. Investigations in the past have not
shown any trend of a turnover in millisecond pulsar spectra (Kramer et
al.~1999). In recent years, however, with the advent of high
sensitivity data acquisition systems, these pulsars are routinely
detectable over a wide range of radio frequencies. Recent work by
Kuniyoshi et al.~2015 shows that a number of millisecond pulsars are
likely to have spectral turnovers at frequencies in the range
0.5--1~GHz. A future application of this model will be to investigate
whether any of these pulsars lie within dense environments 
and use the model to probe the ISM in the vicinity of the pulsar.

\begin{figure}
\centerline{\psfig{file=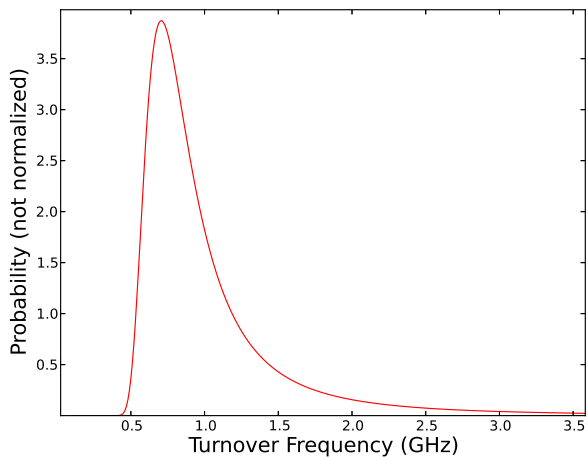,width=9cm}}
\caption{Sample expected probability density function of turnover
  frequencies for a putative line-of-sight to the Galactic center
  obtained using Equation 4 and assuming a distribution of spectral indices for a sample population of pulsars in the Galactic center (see text). \label{fig:PDF}}
\end{figure}

The size and characteristics of the pulsar population in the Galactic
centre (GC) has been a puzzle for astronomers for the past few
years. Several authors have tried to constrain the population of
pulsars in the GC using various techniques (see, e.g., Chennamangalam
\& Lorimer 2014; Macquart \& Kanekar 2015) that try to account for a
number of observational selection effects. The model discussed here might have potential implications on such work. If we adopt values from the
literature for model parameters for a line of sight to the GC, EM = $5
\times 10^{5}~{\rm cm}^{-6}~{\rm pc}$ and $T_{e}= 5000$~K \citep[see,
  e.g.,][]{Pe}, and take a distribution of intrinsic power-law
spectral indices with mean --1.4 and unit standard deviation
\citep{Ba}, using equation~\ref{eq:peak}, it is straightforward to
show that there is a distribution of turnover frequencies that extends
down to a GHz (see Fig.~\ref{fig:PDF}). This distribution suggests
that approximately half of all GC pulsars might exhibit spectral
turnovers at frequencies greater than 1~GHz. Such pulsars would be
harder to detect than previously thought. Recently, there have been
targeted pulsar surveys of the GC at frequencies higher than 1~GHz
(see, e.g., Macquart et al.~2010; Deneva~2010) that should not be
greatly affected by the spectral turnovers. The absence of of any
detected pulsars in these surveys led Chennamangalam \& Lorimer (2014)
to conclude that there are very few of these sources in the GC. The
results found here suggest that the detectability of pulsars in the GC
region may be impacted by spectral turnovers due to the dense
environment. We plan to quantify the impacts of this issue on GC
pulsar population size constraints further in a future paper. 

In summary, we have presented an application of a
    simple free-free absorption model, also proposed by Lewandowski et
    al.~(2015), which is consistent with the turnover in the spectra
of GPS pulsars being caused by propagation through a dense medium. The
results of the thermal absorption model strengthen the claim that 
high-frequency spectral turnovers have their origins in the medium
surrounding the neutron star. We were able to
    determine the most sensible source of absorption for each pulsar
    using an estimate for the mean electron density within the cloud.
More refined measurements of pulsar fluxes, and more examples of GPS
pulsars, are essential to test the model further.

\section*{Acknowledgements}
We thank the referee for the comments that greatly improved the paper. This work was supported by a NASA grant (Proposal number: 74279).

\bibliographystyle{plain}
\bibliography{natbib}
\end{document}